\documentclass{aa}

\usepackage{amsmath}           
\usepackage{dcolumn}           
\usepackage{graphicx}          
\usepackage{txfonts}           
\usepackage{mathrsfs}          
\usepackage{rotating}          
\usepackage{subfigure}         
\usepackage{lscape}            
\usepackage{afterpage}         
\usepackage{xspace}            
\usepackage{natbib}            
\usepackage[version=4]{mhchem} 
\usepackage{hyperref}          
\usepackage{nicefrac}          
\usepackage{amsfonts}          
\usepackage{mathrsfs}          

\providecommand*{\arcsec}{\ensuremath{^{\prime\!\prime}}\xspace}

\providecommand*{\hms}[3]{\ensuremath{{#1}^\mrm{h}{#2}^\mrm{m}{#3}^\mrm{s}}\xspace}
\providecommand*{\dms}[3]{\ensuremath{#1^\circ #2^\prime #3^{\prime\prime}}\xspace}

\newcommand{\mcl}[3]{\multicolumn{#1}{#2}{#3}}
\newcommand{\mrm}[1]{\ensuremath{\mathrm{#1}}}

\newcolumntype{.}{D{.}{.}{-1}}
\newcolumntype{d}[1]{D{.}{.}{#1}}

\begin{document}

	\title{First detection of \ce{NHD} and \ce{ND2} in the interstellar medium}
	\subtitle{Amidogen deuteration in IRAS 16293--2422}

	\author{M.~Melosso\inst{1}
		\and
		L.~Bizzocchi\inst{2}
		\and
		O.~Sipil\"a\inst{2}
		\and
		B.~M.~Giuliano\inst{2}
		\and
		L.~Dore\inst{1}
		\and
		F.~Tamassia\inst{3}
		\and
		M.-A.~Martin-Drumel\inst{4}
		\and
		O.~Pirali\inst{4}\fnmsep\inst{5}
		\and
		E.~Redaelli\inst{2}
		\and
		P.~Caselli\inst{2}
	}
	
	\institute{Dipartimento di Chimica ``Giacomo Ciamician'', Universit\`a di Bologna, via F.~Selmi~2, 40126 Bologna (Italy)\\
		\email{mattia.melosso2@unibo.it}
		\and
		Center for Astrochemical Studies, Max-Planck-Institut f\"ur extraterrestrische Physik, Gie\ss enbachstra\ss e~1, 85748 Garching (Germany)\\
		\email{bizzocchi@mpe.mpg.de}
		\and
		Dipartimento di Chimica Industriale ``Toso Montanari'', Universit\`a di Bologna, viale del Risorgimento~4, 40136 Bologna (Italy)
		\and
		Universit\'e Paris-Saclay, CNRS, Institut des Sciences Mol\'eculaires d'Orsay, 91405 Orsay Cedex (France)
		\and
		SOLEIL Synchrotron, AILES beamline, l'Orme des Merisiers, Saint-Aubin, 91190 Gif-sur-Yvette (France)
	}
	
	\date{Received \today; accepted --}

	\abstract
	{Deuterium fractionation processes in the interstellar medium (ISM) have been shown to be highly efficient in the family of nitrogen hydrides. To date, observations were limited to ammonia (\ce{NH2D}, \ce{NHD2}, \ce{ND3}) and imidogen radical (\ce{ND}) isotopologues.}
	{We want to explore the high frequency windows offered by the \emph{Herschel Space Observatory} to search for deuterated forms of amidogen radical \ce{NH2} and to compare the observations against the predictions of our comprehensive gas-grain chemical model.}
	{Making use of the new molecular spectroscopy data recently obtained at high frequencies for \ce{NHD} and \ce{ND2}, both isotopologues have been searched for in the spectral survey towards the class~0 IRAS 16293-2422, a source in which \ce{NH3}, \ce{NH} and their deuterated variants have been previously detected. We used the observations carried out with HIFI (Heterodyne Instrument for the Far Infrared) in the framework of the key program ``Chemical \emph{Herschel} surveys of star forming regions'' (CHESS).}
	{We report the first detection of interstellar \ce{NHD} and \ce{ND2}. Both species are observed in absorption against the continuum of the protostar. From the analysis of their hyperfine structure, accurate excitation temperature and column density values have been determined. The latter were combined with the column density of the parent species \ce{NH2} to derive the deuterium fractionation in amidogen.
	Our findings point out a high deuteration level of amidogen radical in IRAS 16293-2422, with a deuterium enhancement about one order of magnitude higher than that predicted by earlier astrochemical models. Such a high enhancement can only be reproduced by a gas-grain chemical model if the pre-stellar phase preceding the formation of the protostellar system lasts long, of the order of one million years.
	}
	{The amidogen D/H ratio measured in the low-mass protostar IRAS 16293-2422 is comparable to the one derived for the related species imidogen and much higher than that observed for ammonia. Additional observations of these species will give more insights into the mechanism of ammonia formation and deuteration in the ISM.	We finally indicate the current possibilities to further explore these species at submillimeter wavelengths.}

	
	\maketitle
	%
	
	\section{Introduction}
	\indent\indent
	Despite the low cosmic abundance of deuterium (D/H = $1.5\times10^{-5}$, \citealt{linsky2003atomic}), deuterated molecules can be found in the interstellar medium (ISM) with enormously large amounts
	and are routinely used as proxies for the cold environments typical of the early stages of the star formation process.
	It is well known that deuterium fractionation processes, \textit{i.e.}, the deuterium enrichment in a given molecular species, can occur at very high rate under some specific physical conditions including, above all, a low gas temperature (5--20\,K) and strong \ce{CO} depletion \citep{caselli2012our}.
	Inheriting such conditions from their pre-stellar core phase, Class 0 protostars are the astronomical objects which exhibit the largest molecular deuteration found so far \citep{ceccarelli2014deuterium}. In the last few decades, several deuterium-bearing species (including doubly- and triply-deuterated forms) have been detected in these sources, e.g., water \ce{D2O} \citep{butner2007discovery}, formaldehyde \ce{D2CO} \citep{ceccarelli1998detection}, hydrogen sulfide \ce{D2S} \citep{vastel2003first}, methanol \ce{CD3OH} \citep{parise2004first}, and ammonia \ce{ND3} \citep{roueff2005interstellar,vdTak2002nd3}.
	
	The Class 0 protostar IRAS 16293-2422, located in the nearby $\rho$-Ophiuchi star-forming region (at a distance of $\sim$120\,pc), is considered a specimen for super-deuteration phenomena \citep{ceccarelli2007ps0}. Indeed, besides an exceptionally rich chemistry in its inner and warmer regions \citep{van1995molecular,jorgensen2012detection,fayolle2017protostellar}, this source harbors a large number of interstellar deuterated species in the outer and cold envelope \citep{loinard2001doubly}. 
	Among them, the family of neutral nitrogen hydrides (\ce{NH}, \ce{NH2}, and \ce{NH3}) is well-represented, given that all the deuterated forms of ammonia (\ce{NH2D}, \ce{NHD2}, and \ce{ND3}) and the imidogen radical \ce{ND} \citep{bacmann2010first,bacmann2016stratified} have been detected so far. However, deuterium enrichment has never been observed for the amidogen radical \ce{NH2}, neither in IRAS 16293-2422 nor in any other astronomical sources.
	
	Several factors may be imputed to the current lack of observation of interstellar \ce{NHD} and \ce{ND2}. Until recently, laboratory data were limited to the frequency region below 500\,GHz \citep{kanada1991microwave,kobayashi1997microwave}. At low temperatures, neither species possess any bright transitions in this portion of the spectrum, with $b$-type fundamental transitions lying high into the submillimeter-wave domain (around 770\,GHz for \ce{NHD} and around 527\,GHz for \ce{ND2}).
	Furthermore, the amidogen radical is a textbook floppy molecule, for which frequency extrapolation from low frequency measurements is extremely unreliable.
	Besides the deficiency of laboratory spectral data, ground-based astronomical observations of these light species are hindered by the atmosphere opacity at submillimeter-wavelength. Therefore, even with high-frequency facilities (such as ALMA or SOFIA), the observation of light hydrides remains a challenging task. On-board observations using the \emph{Herschel Space Observatory} mission, during which numerous projects attempted at gaining more insights into the chemistry of the ISM, offer a unique chance to overcome the latter issue.
	Also, now that recent laboratory studies of nitrogen hydrides in their various deuterated forms have been reported up to the terahertz domain \citep{melosso2017terahertz,melosso2019rotational,melosso2019pure,melosso2020nh2d,bizzocchi2020nhd}, confident searches for these species can be undertaken in \emph{Herschel} archival surveys.
	
	In this paper, we report the first detection of \ce{NHD} and \ce{ND2} in the interstellar medium, thanks to HIFI\footnote{Heterodyne Instrument for the Far-Infrared.} observations carried out as part of the CHESS\footnote{Chemical Herschel surveys of star forming regions.} key program (PI: C.~Ceccarelli).
	Both species were observed in absorption against the continuum of the low-luminosity protostar IRAS 16293-2422.
	From the modeling of their hyperfine structures, column densities of \ce{NHD} and \ce{ND2} have been determined and the deuterium fractionation of amidogen radical evaluated.
	
	
	\section{Observations}
	\indent\indent
	The observations analysed in the present paper have been retrieved from the HSA\footnote{ESA \textit{Herschel} Science Archive \url{http://archives.esac.esa.int/hsa/whsa}.}.
	They were collected as part of CHESS key program, whose observational details have been given extensively in \citet{ceccarelli2010herschel}.
	Briefly, the Class 0 protostar IRAS 16293-2422 was observed in March 2010 with the HIFI spectrometer \citep{degraauw2010hifi,roelf2012hifi} for a total of 50\,h.
	The targeted coordinates were $\alpha_{2000} = \hms{16}{23}{22.75}$,  $\delta_{2000} = \dms{-24}{28}{34.2}$\@.
	Seven different HIFI bands were employed to perform a spectral survey covering the frequency range between 480 and 1902\,GHz almost continuously.
	The observations were carried out in double-sideband (DSB) using the wide band spectrometer (WBS), whose resolution is $\sim 1.1$\,MHz.
	
	For the present study we have employed bands 1a (488--552\,GHz), 2b (724--792\,GHz), and 4a (967--1\,042\,GHz), corresponding to the IDs number 1342191499, 1342192332, and 1342191619, respectively.
	HSA provides level~2.5 data products obtained with HIFI pipeline of the \emph{Herschel} interactive processing environment \citep[HIPE,][]{ott2010HIPE}.
	These data generally offer a scientific quality sufficient to perform the line analysis. 
	However, given the weakness of the \ce{NHD} and \ce{ND2} spectral signals, we have also gone through the final processing manually.
	Starting from level~2 products, we have inspected all the data, manually removed bad scans, and exported all the ``good'' 1\,GHz chunks to CLASS/GILDAS\footnote{See GILDAS home page at the URL: \url{http://www.iram.fr/IRAMFR/GILDAS}.} format.
	
	Side-band deconvolution has been performed with the minimization algorithm of \citet{comito2004decon} implemented into CLASS90, and the resulting spectra have been cross-checked with the original level~2.5 products.
	In a few instances, the deconvolution procedure produced a small increase of the noise level compared to the averaged level~2 scans.
	In such cases, we directly use the DSB spectra after verifying that no interfering emission was present in the image band.
	At the later stage, intensive baseline subtraction on H- and V-polarisation spectra has  been applied in the vicinity of each target feature; then, the two set of data have been re-sampled and averaged to obtain the final spectrum.

	
	\section{Analysis}
\begin{figure}[ht!]
		\centering
		\includegraphics[width=\hsize]{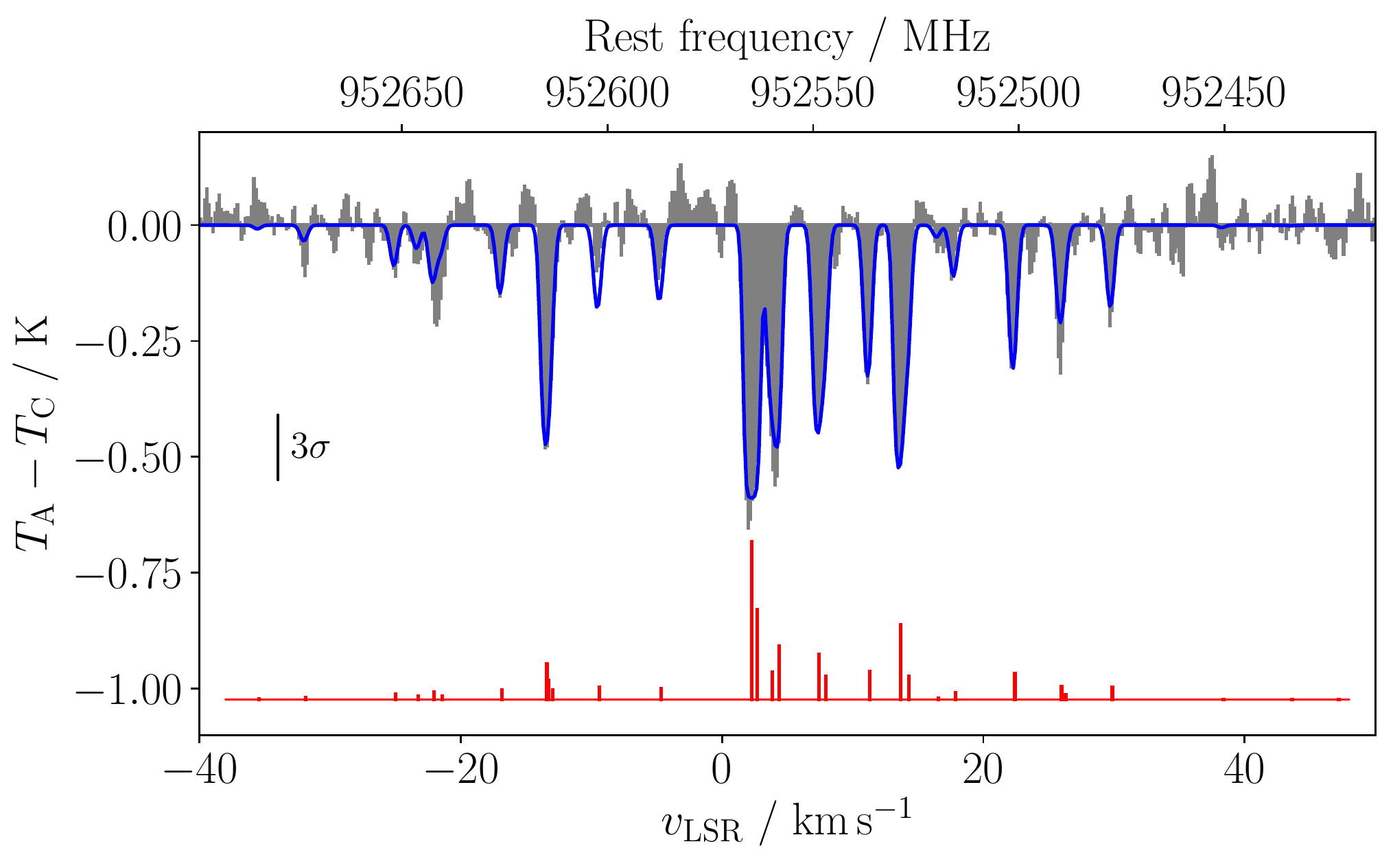}
		\caption{Spectrum of the $J = \tfrac{3}{2} - \tfrac{1}{2}$ component of the $N_{Ka,Kc} = 1_{1,1} - 0_{0,0}$ \textit{o}-\ce{NH2} transition around 952\,GHz observed towards IRAS 16293-2422\@. The filled gray histograms show the observed spectra, while the blue trace is our HFS fit. The red sticks represent the position of the HFS components assuming LTE intensities.}
		\label{fig:nh2}
	\end{figure}

	\indent\indent
	The presence of amidogen radical in IRAS 16293-2422 has already been established by \cite{hily2010nitrogen}, during the early phase of the CHESS key program. 
	The results obtained in the previous study for the parent species \ce{NH2} have been used to test the validity of our data processing and the subsequent spectral analysis.
	Figure~\ref{fig:nh2} shows the strongest $J = \tfrac{3}{2} - \tfrac{1}{2}$ component of the fundamental $N_{Ka,\,Kc} = 1_{1,1} - 0_{0,0}$  transition of $o$-\ce{NH2} around 952\,GHz. Here, most of the HFS components are resolved and the intensity of the absorption lines is well above the $3\sigma$ noise level threshold.
	
	The resulting model of the full HFS (plotted with the solid blue line in Fig.~\ref{fig:nh2}) has been computed using a custom \texttt{Python3} code which exploits a similar approach to that of the \texttt{HFS} method implemented in CLASS.
	Differently from the latter, our code uses the column density ($N$) and the excitation temperature ($T_\mrm{ex}$) as fit parameters, whereas the total opacity of the transition ($\tau$) is regarded as a derived quantity.
	To this aim, the routine makes use of the physical line parameters and implements the computation of the rotational partition function for any given $T_\mrm{ex}$\@.
	The set of the fit parameters also includes the line systemic velocity ($\varv_0$) in the local standard of rest, and the full-width half maximum of the lines (FWHM, $\Delta\varv$).
	Briefly, for a given input the total line opacity is first calculated through:
	\begin{equation} \label{eq:tau}
		\tau = \sqrt{\frac{\ln 2}{16\pi^3}}\:
		\frac{c^3  A_{ul} g_u N}{\nu^3 Q_\mrm{rot}(T_\mrm{ex}) \Delta\varv} 
		\exp\left(\frac{-E_\mrm{low}}{k_\mrm{B} T_\text{ex}}\right) 
		\left[1 - \exp\left(\frac{-h\nu_0}{k_\mrm{B} T_\text{ex}} \right)\right] \,,
	\end{equation}
	
	where $\nu_0$ is the rest frequency of the (unsplit) rotational transition, $c$ is the speed of light, $Q_\mrm{rot}(T_\mrm{ex})$ is the rotational partition function at $T_\mrm{ex}$, $g_u$ is the degeneracy of the upper level, $A_{ul}$ is the Einstein coefficient for spontaneous emission, $E_\mrm{low}$ is the energy of the lower state, $h$ is the Planck constant, and $k_\mrm{B}$ is the Boltzmann constant.
	Then, assuming the local thermodynamic equilibrium (LTE) approximation and Gaussian profiles, the opacity of the $i$-th HFS components is computed as $\tau_i = \tau\,R_i$, where $R_i$ is the corresponding normalised relative intensity.
	The $i$-th component opacity at each velocity channel is then given by:
	\begin{equation} \label{eq:taui}
		\tau_i(\varv) = \tau_i\,\exp\left[-4\ln 2 \left(\frac{\varv - \varv_{0,i}}{\Delta\varv}\right)^2\right] \,,
	\end{equation}
	where $\varv_{0,i}$ is the corresponding velocity position expressed as an offset with respect to the systemic velocity $\varv_0$\@.
	Finally, the continuum-subtracted antenna temperature at a given velocity channel $\varv$ is modeled as
	\begin{equation} \label{eq:RT}
		T_\text{ant}(\varv) = \left[ J_\nu \left(T_\text{ex}\right) - J_\nu \left(T_\text{CMB}\right) - T_\mrm{C} \right]
		\left(1 - \mrm{e}^{-\sum_i\tau_i(\varv)}\right) \,,
	\end{equation}
	where $J_\nu(T)$ is the Rayleigh--Jeans radiation temperature, $T_\mrm{C}$ is the temperature of the continuum and $T_\mrm{CMB}$ is the cosmic background temperature.
	These data are compared to the continuum-subtracted observed antenna temperatures, $T_\mrm{A} - T_\mrm{C}$, in a non-linear least-squares fashion to determine the best fit values of $N$, $T_\mrm{ex}$, $\Delta\varv$, and $\varv_0$\@.
	Whenever possible, all four parameters have been adjusted during the analysis. This sometimes produced strong correlation between them, specially when fitting weak features (e.g., those of \ce{ND2}), for which the relative intensities of the observed HFS components are poorly constrained because of the low signal-to-noise ratio (S/N).
	In these cases, suitable assumption for $T_\mrm{ex}$ and $\Delta\varv$ had to be used and the corresponding parameters were kept fixed in the least-squares fit.
	The single-sideband continuum temperature $T_\mrm{C}$, inserted in the radiative transfer equality of Eq.~\eqref{eq:RT}, has been estimated using the linear formula given by \citet{hily2010nitrogen}, $T_\mrm{C}/\mrm{[K]} =1.10\,\nu - 0.42$, with $\nu$ expressed in THz.
	
	The spectroscopic data for \ce{NH2}, \ce{NHD}, and \ce{ND2} have been taken from the most recent laboratory studies \citep{martin2014nh2,bizzocchi2020nhd,melosso2017terahertz}, and the $Q_\mrm{rot}$ values used in Eq.~\eqref{eq:tau} have been computed at any given temperature by cubic interpolation on a grid of finely spaced entries spanning the 2.7--19\,K interval.
	These were obtained by direct summation on all rotational levels through the \texttt{SPCAT} spectral tool
	\citep{pickett1991}, with the $ortho$ and $para$ species of the symmetric isotopologues \ce{NH2} and \ce{ND2} treated separately.
	A brief explanation of the energy levels structure of \ce{NHD} and \ce{ND2} is given in Appendix~\ref{app:levels}, while the complete list of HFS components used for the analysis of each transition is reported in Appendix~\ref{app:list}. The partition function values of both species computed at temperatures between 2.725 and 300\,K are given in Appendix~\ref{app:partit}.
	
	
	\section{Results}
	\indent\indent
	The line parameters obtained from the analysis of the observed transitions of \ce{NH2}, \ce{NHD}, and \ce{ND2} are collected in Table~\ref{tab:obsres}.
	In the first two rows, the results obtained for the fine-structure line of $o$-\ce{NH2} detected previously by \citet{hily2010nitrogen} are reported.
	Our fitted $N$ and $T_\mrm{ex}$ values compare very well with those obtained from the CLASS \texttt{HFS} method employed in that study. 
	Our weighted average of the column density is $(5.4\pm 0.4)\times 10^{13}$\,cm$^{-2}$, which is consistent within the uncertainty with the value $(4.4\pm 0.7)\times 10^{13}$\,cm$^{-2}$ obtained previously.
	The excitation temperatures determined for the two components are $1\sigma$ coincident ($8.44\pm 0.12$\,K the average), and are also close to the ones obtained by \citet{hily2010nitrogen} (8.5\,K for the $J = \tfrac{3}{2} - \tfrac{1}{2}$ and 9.5\,K for $J = \tfrac{1}{2} - \tfrac{1}{2}$).
	
\begin{figure}[ht!]
		\centering
		\includegraphics[width=\hsize]{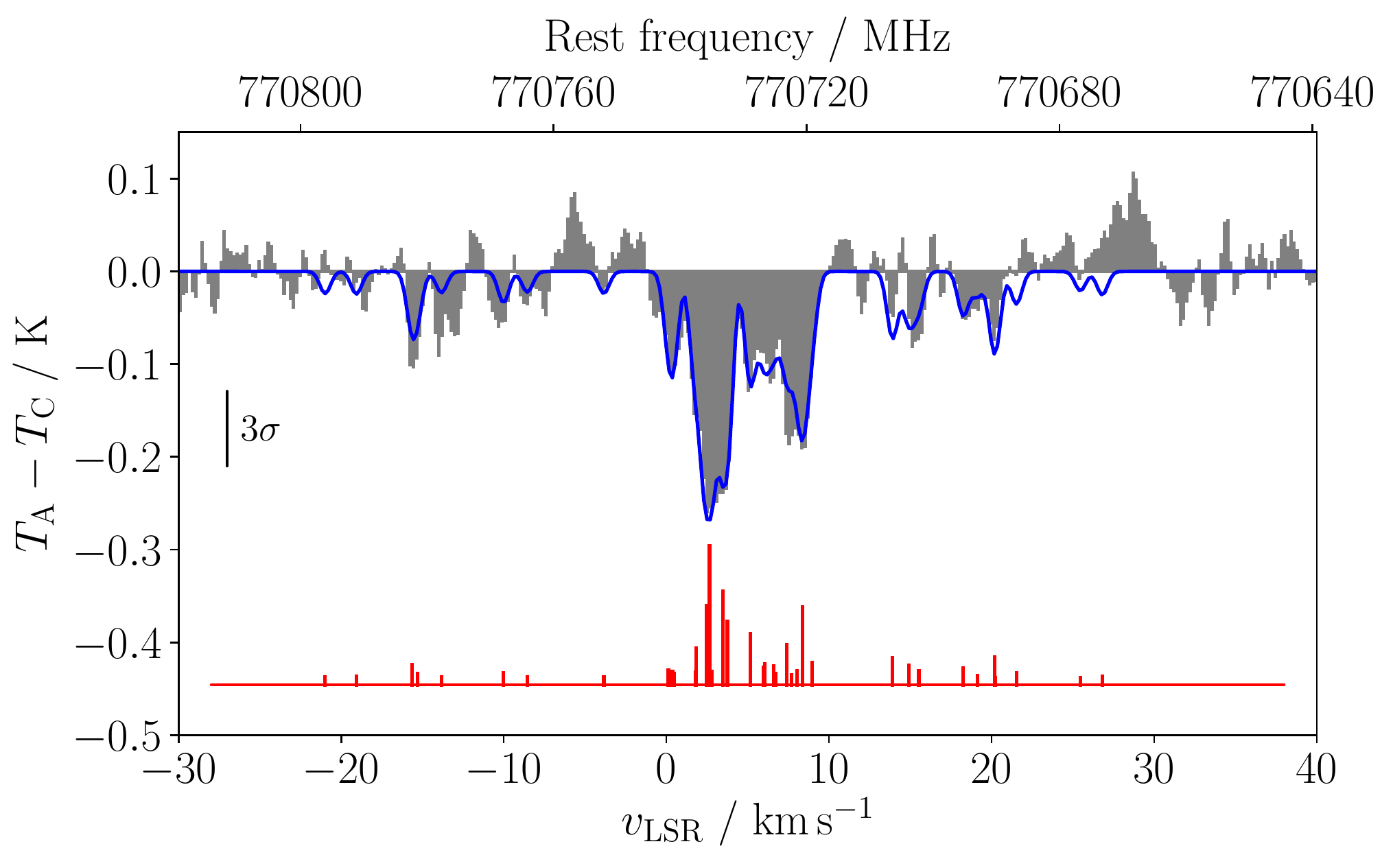}
		\includegraphics[width=\hsize]{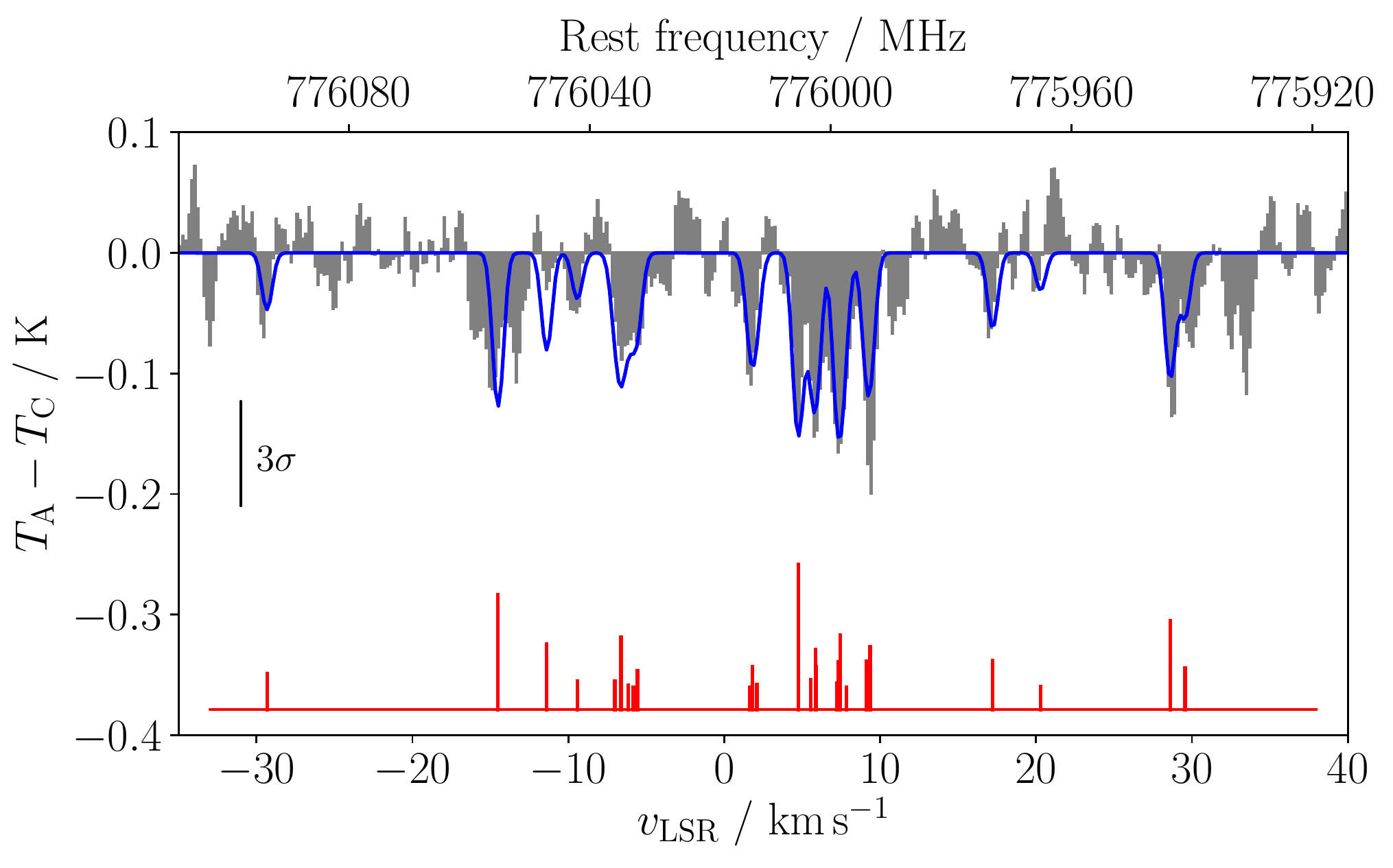}
		\caption{Spectra of the fine-components of the $N_{Ka,Kc} = 1_{1,1} \leftarrow 0_{0,0}$ \ce{NHD} transition observed towards IRAS 16293-2422\@. \textit{Upper panel:} $J = \tfrac{3}{2} - \tfrac{1}{2}$ component at \textit{ca.}~770\,GHz. \textit{Lower panel:} $J = \tfrac{1}{2} - \tfrac{1}{2}$ component at \textit{ca.}~776\,GHz.  The filled gray histograms show the observed spectra, while the blue trace is our HFS fit. The red sticks represent the position of the HFS components assuming LTE intensities.}
		\label{fig:nhd}
	\end{figure}

	In the same spectral survey, we detect for the first time two groups of hyperfine transitions belonging to the singly-deuterated form of amidogen radical, \ce{NHD}.
	Figure~\ref{fig:nhd} shows the two fine-structure components of the fundamental $N_{Ka,Kc} = 1_{1,1} \leftarrow 0_{0,0}$ transition of \ce{NHD}, both observed in absorption around 770 and 776\,GHz in the 2b HIFI band.
	
	\begin{figure}[t]
		\centering
		\includegraphics[width=\hsize]{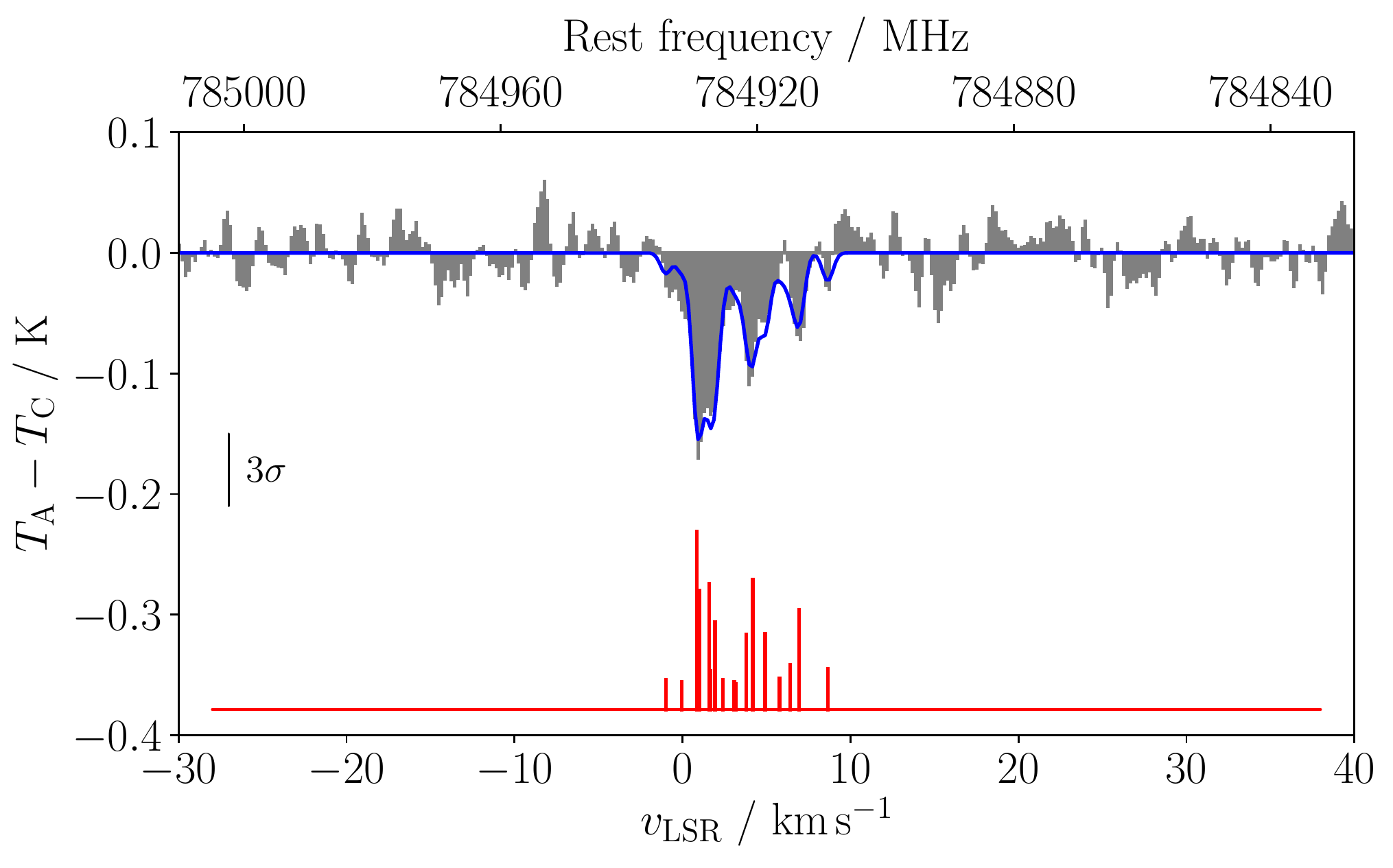}
		\includegraphics[width=\hsize]{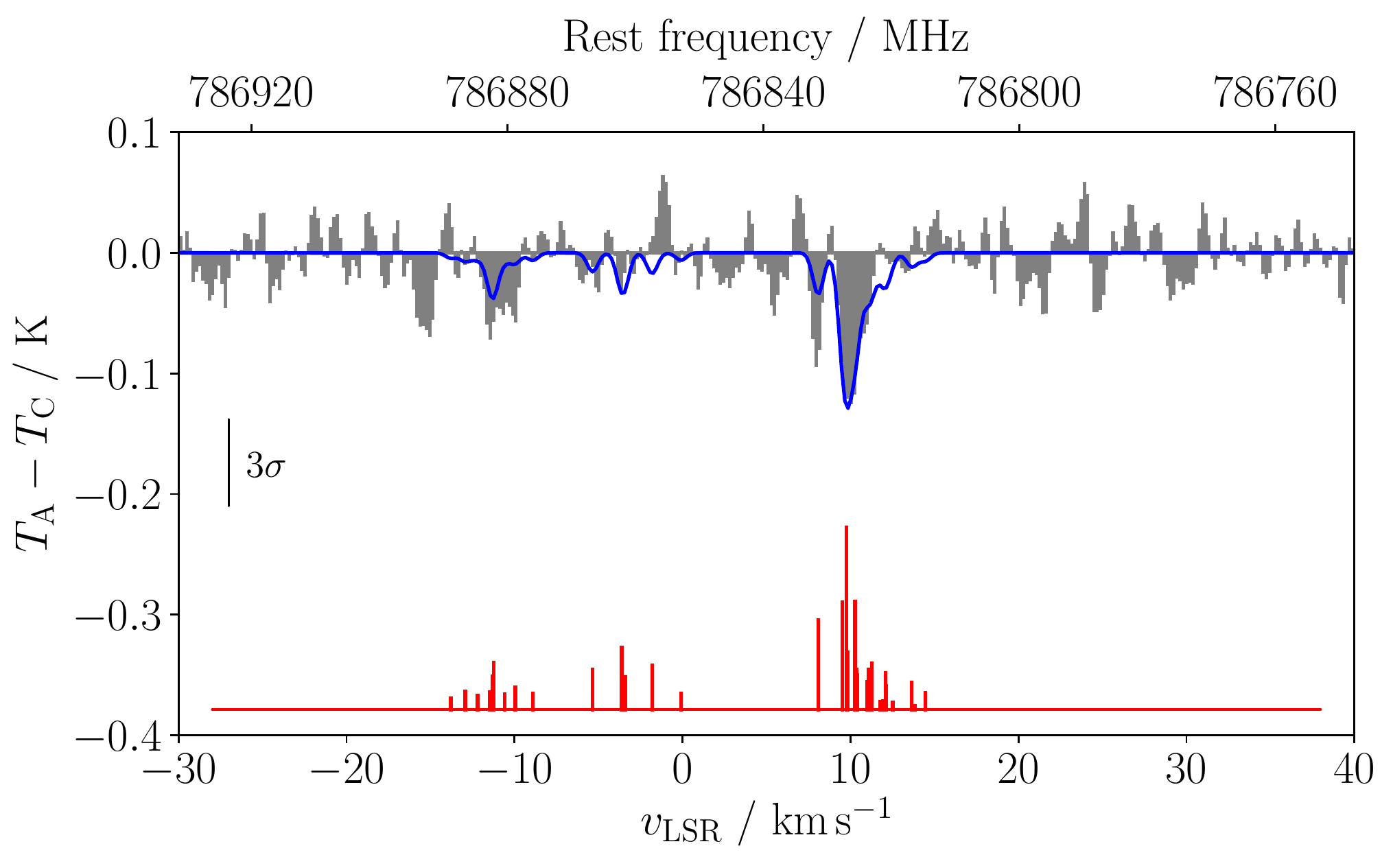}
		\caption{Spectra of the fine-structure components of the $N_{Ka,Kc} = 2_{1,2} \leftarrow 1_{0,1}$ $p$-\ce{ND2} transition observed towards IRAS 16293-2422\@.
			\textit{Upper panel:} $J = \tfrac{5}{2} - \tfrac{3}{2}$ component at ca.~785\,GHz. 
			\textit{Lower panel:} $J = \tfrac{3}{2} - \tfrac{1}{2}$ component at ca.~787\,GHz.  
			The filled grey histograms show the observed spectra, while the blue trace is our HFS fit. 
			The red sticks represent the position of the HFS components assuming LTE intensities.}
		\label{fig:nd2_21}
	\end{figure}
	
	\begin{table*}[t]
		\centering
		\caption{Summary of the transitions observed for \ce{NH2}, \ce{NHD}, and \ce{ND2} towards IRAS 16293-2422\@.
			\label{tab:obsres}}
		\begin{tabular}{l c d{4} c d{2} d{3} d{6} d{5} d{7} d{3}}
			\hline\hline \\[-1.5ex]
			Species & Transition & \mcl{1}{c}{Frequency\,\tablefootmark{a}} 
			& \mcl{1}{c}{HPBW\,\tablefootmark{b}}
			& \mcl{1}{c}{$T_\mrm{C}$} 
			& \mcl{1}{c}{$T_\mrm{ex}$} 
			& \mcl{1}{c}{$\varv_\mrm{LSR}$}  
			& \mcl{1}{c}{$\Delta\varv$\,\tablefootmark{c}} 
			& \mcl{1}{c}{$N$}         
			& \mcl{1}{c}{$\tau$\,\tablefootmark{d}} \\[0.5ex]
			& $N_{K_a,Kc},J$  & \mcl{1}{c}{[GHz]}     
			& \mcl{1}{c}{[arcsec]}
			& \mcl{1}{c}{[K]}          
			& \mcl{1}{c}{[K]}          
			& \mcl{1}{c}{[km\,s$^{-1}$]}
			& \mcl{1}{c}{[km\,s$^{-1}$]}
			& \mcl{1}{c}{[$10^{13}$\,cm$^{-2}$]} 
			& \mcl{1}{c}{}       \\[0.5ex]
			\hline \\[-1.5ex]
			$o$-\ce{NH2} & $1_{1,1},\tfrac{3}{2} - 0_{0,0},\tfrac{1}{2}$ &  952.5722  &  22.4  &  0.81  &  8.6(1)  &  4.096(9)  &  0.67(8)  &  5.3(3)   & 20.3(13)  \\[0.5ex]
			& $1_{1,1},\tfrac{1}{2} - 0_{0,0},\tfrac{1}{2}$ &  959.5043  &  22.1  &  0.82  &  8.8(3)  &  4.077(9)  &  0.66(8)  &  6.8(9)   & 13.0(12)  \\[1.0ex]
			\ce{NHD}     & $1_{1,1},\tfrac{3}{2} - 0_{0,0},\tfrac{1}{2}$ &  770.7422  &  27.5  &  0.55  &  7.4(1)  &  4.19(3)   &  0.78(11) &  4.4(6)   & 10.3(15)  \\[0.5ex]
			& $1_{1,1},\tfrac{1}{2} - 0_{0,0},\tfrac{1}{2}$ &  776.0177  &  27.3  &  0.56  &  7.4     &  4.10(3)   &  0.78     &  5.0(3)   &  5.8(5)   \\[1.0ex]
			$p$-\ce{ND2} & $2_{1,2},\tfrac{5}{2} - 1_{0,1},\tfrac{3}{2}$ &  784.9317  &  27.0  &  0.57  &  7.4     &  4.23(4)   &  0.78     &  0.23(1)  &  2.4(2)   \\[0.5ex]
			& $2_{1,2},\tfrac{3}{2} - 1_{0,1},\tfrac{1}{2}$ &  786.8526  &  26.9  &  0.57  &  7.4     &  4.31(6)   &  0.78     &  0.27(2)  &  1.5(2)   \\[0.5ex]
			$o$-\ce{ND2} & $1_{1,1},\tfrac{3}{2} - 0_{0,0},\tfrac{1}{2}$ &  527.1808  &  40.2  &  0.20  &  4.5     &  3.98(7)   &  0.83     &  0.66(8)  &  2.0(3)   \\[0.5ex]
			\hline \\[-1.5ex]
		\end{tabular}
		\tablefoot{Numbers in parentheses refer to 1$\sigma$ uncertainties expressed in units of the last quoted digit. \tablefoottext{a}{Rest frequency corresponding to the hypothetically unsplit fine-structure transition.} \tablefoottext{b}{Half-power beam width of the observations.} \tablefoottext{c}{A conservative uncertainty of 0.25\,MHz produced by the HIPE pipeline is summed in quadrature.} \tablefoottext{d}{Derived through Eq.~\eqref{eq:tau}.}}
	\end{table*}
	
	Since the $J = \tfrac{3}{2}-\tfrac{1}{2}$ fine-structure line is detected at ca.~10$\sigma$ level, we could adjust all the four parameters in the model.
	The $J = \tfrac{1}{2}-\tfrac{1}{2}$ component is about twice weaker and its detection is only at ca.~5$\sigma$ level.
	In this case, the noise prevented to adjust $T_\mrm{ex}$ and $\Delta\varv$, which were then fixed at the values obtained for the strongest component of the fine-structure doublet.
	For \ce{NHD}, we found an excitation temperature of $7.4 \pm 0.1$\,K, slightly lower than that of \ce{NH2}. 
	This small difference is likely to be produced by the bigger \textit{Herschel} beam size in the 2b band ($\sim 27$\,\arcsec) compared to that in the 4a band ($\sim 22$\,\arcsec).
	Indeed, a larger region is sampled for the \ce{NHD} lines, and the contribution of the inner warmer gas is expected to be slightly smaller.
	The average value of the column density is $N = (4.7\pm 0.7)\times 10^{13}$\,cm$^{-2}$\@.
	
\begin{figure}[ht!]
		\centering
		\includegraphics[width=\hsize]{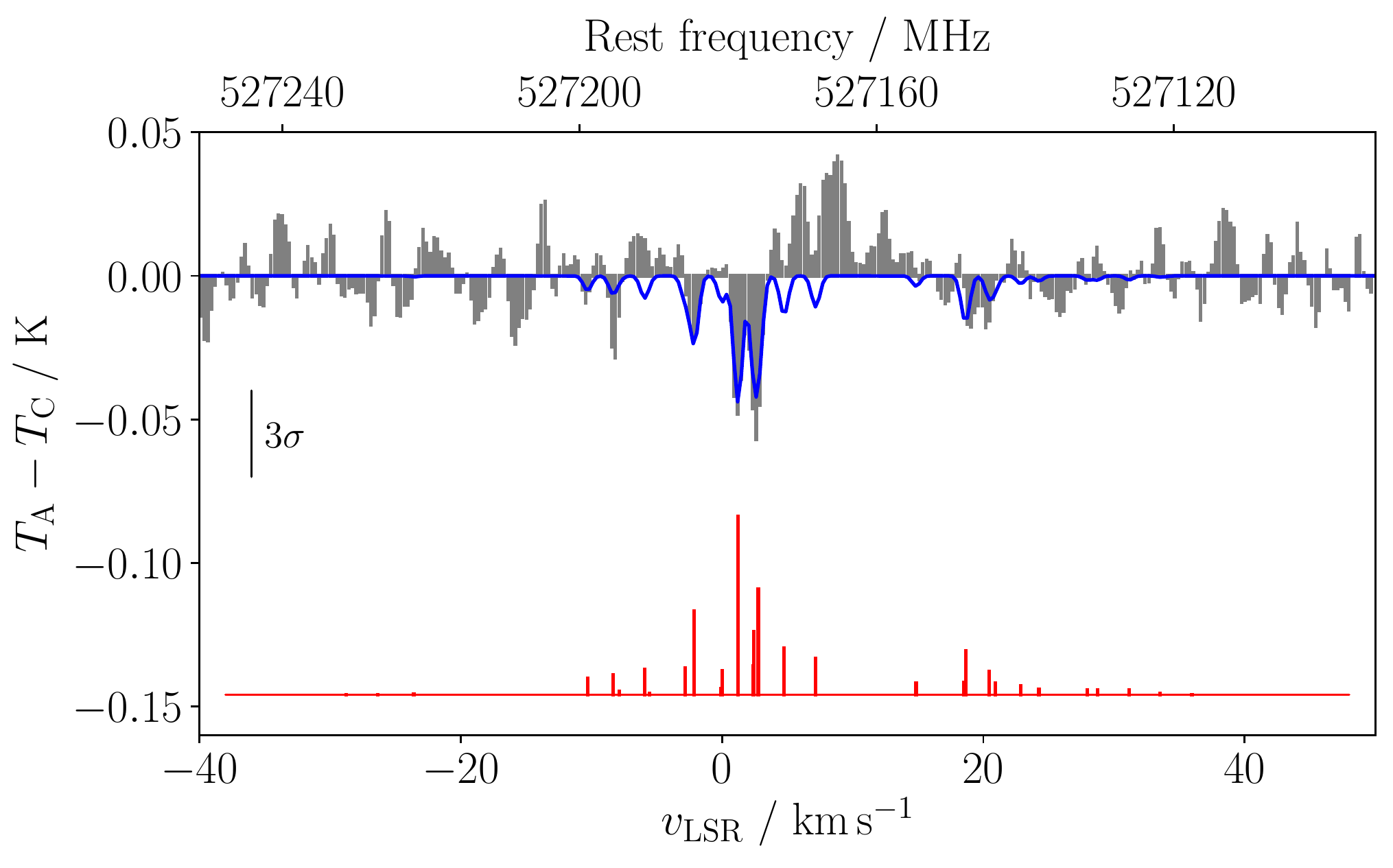}
		\caption{Spectrum of the $J = \tfrac{3}{2} - \tfrac{1}{2}$ component of the $N_{Ka,Kc} = 1_{1,1} - 0_{0,0}$ \textit{o}-\ce{ND2} transition around 527\,GHz observed towards IRAS 16293-2422\@. 
			The filled gray histograms show the observed spectra, while the blue trace is our HFS fit. 
			The red sticks represent the position of the HFS components assuming LTE intensities.}
		\label{fig:nd2_10}
	\end{figure}

	Compared to the parent isotopologue, the doubly deuterated form of amidogen has a denser spectrum in the sub-millimeter region, thus both the $N_{Ka,\,Kc} = 1_{1,1} \leftarrow 0_{0,0}$ and $N_{Ka,\,Kc} = 2_{1,2} \leftarrow 1_{0,1}$ rotational lines fall within the coverage of the CHESS spectral survey.
	A clear detection has been obtained in the 2b band for both $J = \tfrac{5}{2}-\tfrac{3}{2}$ and $J = \tfrac{3}{2}-\tfrac{1}{2}$ fine-structure components of the $N_{Ka,\,Kc} = 2_{1,2} \leftarrow 1_{0,1}$ transitions of the \textit{para} species. They are shown in Fig.~\ref{fig:nd2_21}.
	The $N_{Ka,\,Kc} = 1_{1,1} \leftarrow 0_{0,0}$, $J = \tfrac{3}{2}-\tfrac{1}{2}$ of the \textit{ortho} species, illustrated in Fig.~\ref{fig:nd2_10}, is located at 527\,GHz.
	
	In this spectral region, covered by the HIFI 1a band, the continuum emission of the source is only 200\,mK; hence, the resulting absorption feature is rather faint and the detection is only marginal, reaching at most $5\sigma$\@.
	The other fine-structure component (intrinsically less intense) is just below the noise level and not detected.
	Given the weakness of all the three components observed for \ce{ND2}, their analysis had to be carried out adopting suitable assumptions for $T_\mrm{ex}$ and $\Delta\varv$, as these parameters could not be set free in the least-squares fits.
	Given the similarity of the beam size ($\sim 27$\,\arcsec), for the $p$-\ce{ND2} $N_{Ka,\,Kc} = 2_{1,2} \leftarrow 1_{0,1}$ lines we have assumed the same excitation temperature determined for \ce{NHD}.
	A smaller value is expected to hold for the $o$-\ce{ND2} $N_{Ka,\,Kc} = 1_{1,1} \leftarrow 0_{0,0}$ line, as the \textit{Herschel} beam HPBW at this frequency is as broad as 40\,\arcsec, thus sampling a comparatively large volume of cold gas.
	For this transition we have fixed $T_\mrm{ex} = 4.5$\,K, a value determined for \ce{ND} in the same source \citep{bacmann2010first} through the analysis of its $N=1-0$ fine-structure doublet centered at 534\,GHz.
	In fact, this value allows for a satisfactory fit of the absorption $o$-\ce{ND2} feature, whereas
	for higher excitation temperatures (e.g., already at 5.5\,K) the LTE model predicts the line to be observable in emission.
	Constraints for the FWHM line width $\Delta\varv$ have been derived from the values determined for \ce{NH2} after deconvolution from the WBS spectrometer resolution (1.1\,MHz).
	The column densities derived for $p$-\ce{ND2} using two components are consistent within $2\sigma$ and the weighted average value is $N = (0.241\pm 0.014)\times 10^{13}$\,cm$^{-2}$\@.
	The observed \textit{ortho}-to-\textit{para} ratio is $2.7\pm 0.4$\@.
	By considering (i) the energy difference of 10.77\,cm$^{-1}$ ($E/k\approx 15.5$\,K) between the lowest \textit{para} and \textit{ortho} rotational levels, and (ii) the spin statistical weights, this ratio corresponds to an equilibrium temperature of $9\pm 1$\,K\@.
	Due to the different beam sizes used to observe the $N_{Ka,\,Kc} = 1_{1,1} \leftarrow 0_{0,0}$, and $2_{1,2} \leftarrow 1_{0,1}$ transitions, the $p$-\ce{ND2} is preferentially sampled in the inner warmer region; therefore, this value is likely to be slightly overestimated.
	
	For the same reason, the evaluation of the abundance ratios between amidogen isotopologues requires some careful considerations.
	From the averaged $N$ values we obtained $[\ce{NHD}]/[\ce{NH2}] = 0.87\pm 0.14$, hinting to a very high deuteration level for amidogen, lying in the 70-100\% range. We consider this results as robust.
	The HPBW of the observations differs by only 25\%, thus the source area sampled by the antenna is similar for both \ce{NHD} and \ce{NH2} lines.
	Furthermore, the energy difference between \textit{ortho} and \textit{para} species of \ce{NH2} is 21.11\,cm$^{-1}$ ($E/k_\mrm{B}\approx 30.4$\,K), and the contribution of the unobserved $p$-\ce{NH2} is expected to be less than 5\% at 10\,K\@.
	Finally, since the used transitions have their lower level located at ``zero'' energy, the product of two exponential terms in Eq.~\eqref{eq:tau} tends to 1 for small $h\nu_0/k_\mrm{B}T_\mrm{ex}$, and the relation between $N$ and $\tau$ depends only mildly on the temperature.
	Hence, the retrieved column density values are not much affected by the inaccuracies in the corresponding $T_\mrm{ex}$ determination.
	
	The derivation of sound values of the isotopic ratios involving \ce{ND2} is less straightforward, as various approaches can be adopted to estimate the overall \ce{ND2} column density from the available observational results.
	By assuming that the two spin species are in thermal equilibrium, total $N$ values of $0.70\times 10^{13}$\,cm$^{-2}$ and $0.98\times 10^{13}$\,cm$^{-2}$ can be obtained from the $o$-\ce{ND2} and $p$-\ce{ND2} results, respectively.
	A direct sum of the \textit{ortho} and \textit{para} column density with no assumptions gives instead $0.90\times 10^{13}$\,cm$^{-2}$\@.
	Based on the resulting 30\% discrepancies, we conservatively quote
	$N = (0.9\pm 0.3)\times 10^{13}$\,cm$^{-2}$ as the total column density of \ce{ND2}\@.
	This yields $[\ce{ND2}]/[\ce{NHD}] = 0.19\pm 0.07$ and $[\ce{ND2}]/[\ce{NH2}] = 0.17\pm 0.06$, that is, deuteration levels in the 10--30\% range.
	The amidogen D/H ratio measured in the low-mass protostar IRAS16293 is very high and is comparable to the  one derived for the related species imidogen ([\ce{ND}]/[\ce{NH}]$=30-70$\%, \citealt{bacmann2010first}), while ammonia shows a lower level ([\ce{NH2D}]/[\ce{NH3}]$\sim 10$\%, \citealt{van1995molecular}).
	The \ce{ND2} species is very abundant, yielding a remarkably high deuteration ratio, even higher than the one found for methanol ([\ce{CHD2OH}]/[\ce{CH3OH}]$=6$\%, \citealt{parise2004first}), a species for which the triply deuterated species was also detected in this source.
	
	
	\section{Chemical simulation} \label{sec:simul}
	\indent\indent
	A steady-state model of cold and \ce{CO}-depleted clouds was published by \citet{roueff2005interstellar}, but their predictions on deuterium enhancement, i.e., 10\% for \ce{NHD} and below 1\% for \ce{ND2}, are significantly underestimated with respect to our findings, which are 87\% and 18\% for \ce{NHD} and \ce{ND2}, respectively.
	
\begin{figure*}[t]
		\centering
		\includegraphics[width=2.0\columnwidth]{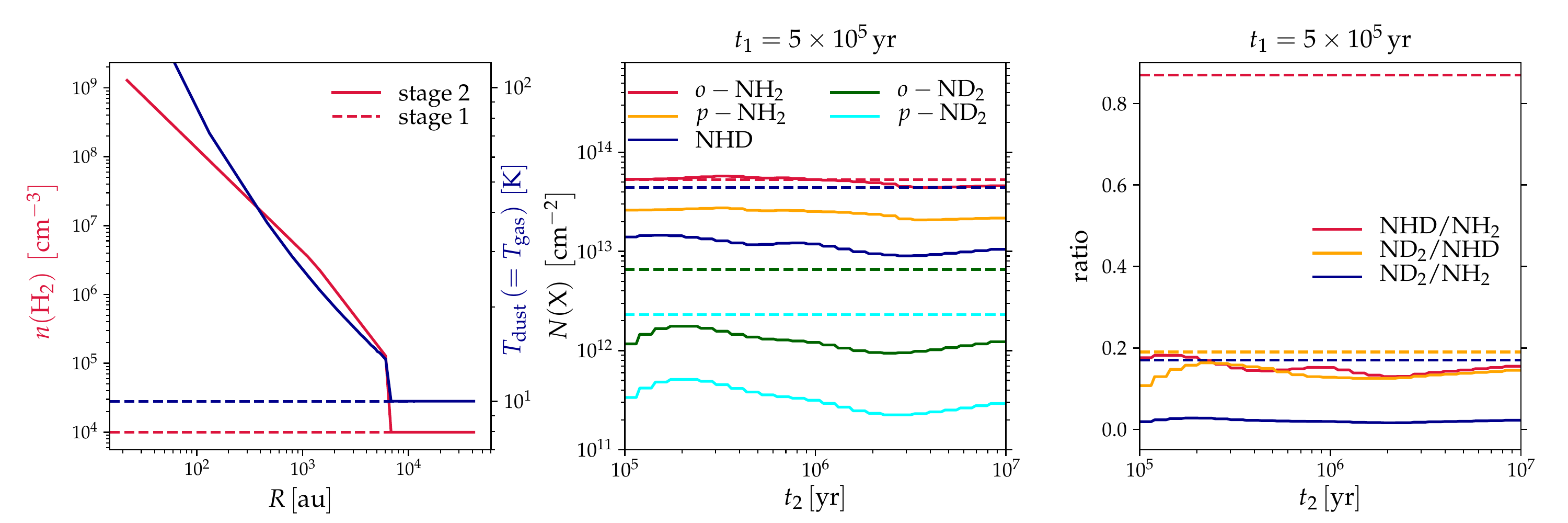}
		\caption{{\sl Left panel:} Density (red) and temperature (blue) structure in the IRAS16293 source model in stages~1~and~2 (see Sect.~\ref{sec:simul} for details). 
			{\sl Middle panel:} Modelled beam-convolved molecular column densities (solid lines, labeled in the plot) as functions of time in stage~2. Dashed lines represent the observed values, omitting error bars for clarity. 
			{\sl Right panel:} Modelled D/H ratios (solid lines, labeled in the plot) as functions of time in stage~2. Here, the \ce{NH2} and \ce{ND2} abundances have been summed over the ortho and para forms. Dashed lines represent the observed values, for which the error bars have again been omitted.
			\label{fig:IRAS16293_bestfit}}
	\end{figure*}

	In an effort to reproduce the observed tendencies of the D/H and spin-state abundance ratios, we have carried out new chemical simulations of (deuterated) amidogen using a model of IRAS16293-2422\@.
	We followed the approach described in \citet{Brunken14} and \citet{Harju17b}, where more details can be found.
	Here, we provide only a brief description of the simulations. For the source model we adopted the one used in the above references, where the physical model from \citet{Crimier10}, which represents the protostellar core, was extended with a layer of gas representing the molecular cloud that the protostellar system lies in (with $n({\rm H_2}) = 10^4 \, \rm cm^{-3}$ and $T_{\rm dust} = T_{\rm gas} = 10\,\rm K$). 
	The source model is illustrated in Fig.\,\ref{fig:IRAS16293_bestfit}. 
	We considered a two-stage model, in which the chemical evolution was first calculated over time $t_1$ in physical conditions corresponding to the core extension (stage~1), and then the protostellar core model including the low-density extension was run over time $t_2$ (stage~2) using the initial chemical abundances obtained from the stage~1. 
	We explored a range of $t_1$ and $t_2$ values to search for the best fit to the observed column densities. 
	
	The simulation results depend on how the chemical reactions related to (deuterated) amidogen formation are assumed to proceed. 
	Here we used our gas-grain chemical code \citep{Sipila19a} and ran a parameter-space exploration, testing the effects of multilayer (three-phase) ice chemistry \citep{Sipila16} and direct proton/deuteron hop \citep{Sipila19b} as opposed to full scrambling in proton-donation reactions \citep{Sipila15b} on the various column densities. 
	As an example of the results, Fig.~\ref{fig:IRAS16293_bestfit} shows a comparison of the beam-convolved modelled and observed column densities and column density ratios using the multi-layer ice model and full scrambling for proton-donation reactions. 
	The chemical model reproduces very well the column density of $o$-\ce{NH2} and the \ce{ND2}/\ce{NHD} ratio, but fails to reproduce the \ce{NHD}/\ce{NH2} ratio as well as the \ce{NHD} and (\textit{ortho} and \textit{para}) \ce{ND2} column densities. 
	Other parameter combinations reproduce instead very well the \ce{NHD} and (\textit{ortho} and \textit{para}) \ce{ND2} column densities, but strongly overpredict the $o$-\ce{NH2} column density. 
	None of the parameter combinations we explored lead to the observed level for the \ce{NHD}/\ce{NH2} ratio. 
	The chemical model predicts that most of the \ce{NHD} column originates near the edge of the protostellar core, where the temperature in the source model ($\sim$14\,K at the very edge, rising inward; cf. Fig.\,\ref{fig:IRAS16293_bestfit}) inhibits \ce{NHD} production with respect to \ce{NH2}. 
	Indeed, a \ce{NHD}/\ce{NH2} ratio of unity or above can be produced by the chemical model if the gas is sufficiently dense ($n({\rm H_2}) \gtrsim 10^5 \, \rm cm^{-3}$) and cold ($T \sim 10 \, \rm K$), suggesting that our two-phase approach to the core modelling is inadequate for the modelling of amidogen chemistry, and that a substantial amount of time (of the order of $10^6 \, \rm yr$) is required in the pre-stellar phase for the \ce{NHD}/\ce{NH2} ratio to increase to the observed level.
	We stress that in the present work, we did not carry out radiative transfer simulations of the amidogen absorption lines, and therefore the comparison of the modelled column densities against the observed ones does not take into account any excitation effects. We will present an in-depth comparison of the model versus the observations in an upcoming work, where we will also carry out line simulations.
	
	
	\section{Discussion and conclusions}
	\indent\indent
	This work presents the very first detection of both deuterated forms of amidogen radical in the ISM. From the observations, column densities and deuterium fractionation values have been derived allowing for a direct comparison with the predictions obtained from chemical networks.
	We have run a parameter-space exploration using our gas-grain chemical code, implementing a two-stage approach to the dynamical evolution of the source.
	The observed abundances of all \ce{NH2}, \ce{NHD}, and \ce{ND2} isotopic variants could not be simultaneously reproduced with a single set of modelling parameters, thus indicating that the two-stage approach to the core modelling is inadequate for simulating amidogen isotopic chemistry.
	More insights on this topic would come from additional observations of deuterated amidogen isotopologues.
	They are, however, partially hampered by the atmospheric opacity. In fact, The most intense transition of \ce{NHD}, located at 770\,GHz, is not accessible with the ground-based facilities because of nearby strong water absorption lines.
	Hence, \ce{NHD} observations can rely mainly on the $a$-type $N_{Ka,\,Kc} = 1_{\,0,\,1} \leftarrow 0_{\,0,\,0}$ transition around 413\,GHz. Luckily, spectral coverage at this frequency is supported by the ALMA 8 Band and the new nFLASH receiver of APEX.
	As for \ce{ND2}, the transitions detected in this work fall in spectral windows accessible either with the incoming band~1 of the 4GREAT receiver on board SOFIA (490--635\,GHz) or ALMA 10 Band (787--950\,GHz).
	
	The conclusions of the present work can be summarized as follows:
	\begin{enumerate}
		\item Thanks to the recently obtained laboratory data, two deuterated forms of amidogen radical, namely \ce{NHD} and \ce{ND2}, have been identified for the first time in the ISM.
		The analysis of hyperfine resolved transitions, observed in absorption against the continuum of the Class 0 protostar IRAS 16293-2422, allowed the accurate determination of column densities and excitation temperatures.
		\item The deuterium fractionation of amidogen results to be 70-100\% for \ce{NHD} and 10--30\% for \ce{ND2}. Our derived values differ significantly, being larger by about one order of magnitude, from those predicted by earlier astrochemical models. 
		\item The low temperature and the high degree of deuteration observed suggest that amidogen, similarly to other nitrogen-containing species, is less depleted and subsists longer in the gas-phase, thus leading to high abundances of \ce{NHD} and \ce{ND2}.
		\item The amidogen D/H and spin-state abundance ratios in IRAS 16293-2422 cannot be reproduced with our gas-grain chemical model unless the pre-stellar stage lasts long, of the order of one million years.
		\item ALMA represents the best candidate to perform future observations of \ce{NHD} and \ce{ND2}, not only in IRAS 16293-2422, but also in other sources where deuterated ammonia has been detected. More observations would enable to make a step towards a thorough understanding of ammonia formation and deuteration in the ISM.
	\end{enumerate}
	
	
	\begin{acknowledgements}
		This study was supported by Bologna University (RFO funds) and by MIUR (Project PRIN 2015:
		STARS in the CAOS, Grant Number 2015F59J3R).
		The work at SOLEIL was supported by the Programme National ``Physique et Chimie du Milieu Interstellaire'' (PCMI) of CNRS/INSU with INC/INP co-funded by CEA and CNES. 
	\end{acknowledgements}
	
	
	\bibliographystyle{aa}
	\bibliography{nhd_iras16293}
	
	
	\begin{appendix}
		
		\section{Energy levels scheme}\label{app:levels}
		
		\subsection{Singly-deuterated amidogen radical}
		\indent\indent
		The manifold of rotational energy levels is very complex in \ce{NHD} and its spectrum is characterized by many interactions. First, \ce{NHD} is a light asymmetric rotor far from the prolate limit ($\kappa=-0.66$) whose electric dipole moment lies in the $ab$ principal symmetry plane, with components $\mu_a = 0.67$\,D and $\mu_b = 1.69$\,D \citep{brown1979determination}. Because of the large values of the rotational constants, the room temperature spectrum is quite sparse and peaks in the far-infrared (FIR) region.
		
		\ce{NHD} is an open-shell molecule with a $\tilde{X}\,^2A^{\prime\prime}$ ground electronic state.
		Due to the presence of an unpaired electron, the electronic spin ($\mathbf{S}=\nicefrac{1}{2}$) couples with the rotational angular momentum $\mathbf{N}$, thus splitting each rotational level $N_{K_a,\,K_c}$ into two fine-structure sub-levels. The fine-structure levels have either $J = N + \nicefrac{1}{2}$ or $J = N - \nicefrac{1}{2}$ quantum numbers, with the exception of the $N = 0$ level, for which only the $J = \nicefrac{1}{2}$ level exists. The total angular momentum $\mathbf{J}$ further couples with each nuclear spin ($I_\mrm{N}$ and $I_\mrm{D}=1$, $I_\mrm{H}=\nicefrac{1}{2}$), giving rise to the so-called hyperfine structure (HFS). 
		The angular momentum coupling scheme adopted takes into account the magnitude of the interaction (from strongest to weakest):
		\begin{subequations} \label{eq:coup}
			\begin{eqnarray} \label{eq:subcoup}
				&\mathbf{J}    = &\mathbf{N}   + \mathbf{S}         \,, \\
				&\mathbf{F}_1  = &\mathbf{J}   + \mathbf{I}_\mrm{N} \,, \\
				&\mathbf{F}_2  = &\mathbf{F}_1 + \mathbf{I}_\mrm{H} \,, \\
				&\mathbf{F}    = &\mathbf{F}_2 + \mathbf{I}_\mrm{D} \,.
			\end{eqnarray}
		\end{subequations}
		A more detailed accounts of the \ce{NHD} spectroscopy can be found in \citet{bizzocchi2020nhd}.
		
		\subsection{Doubly-deuterated amidogen radical}
		\indent\indent
		The higher symmetry of \ce{ND2} with respect to \ce{NHD} requires additional considerations in the description of the energy levels manifold. \ce{ND2} belongs to the $C_{2v}$ symmetry point group and a $\pi$-rotation along the $b$ axis exchanges the two identical deuterium nuclei. Since deuterium is a boson, the total wave-function of \ce{ND2} must be symmetric upon the permutation of the two particles.
		Given that the electronic ground state is antisymmetric ($\tilde{X}\,^2B_1$) and the vibrational ground state is symmetric, deuterium nuclear spin functions ($I_\mrm{D,\,tot}=I_\mrm{D_1}+I_\mrm{D_2}$) can only combine with rotational levels of opposite symmetry. That is, symmetric rotational levels ($K_a+K_c$ = even) combine with antisymmetric deuterium spin functions, with $I_\mrm{D,\,tot}=1$ (\emph{ortho} states), while antisymmetric rotational levels ($K_a+K_c$ = odd) combine with the spin functions $I_\mrm{D,\,tot}=0,\,2$ (\emph{para} states).
		Only transitions within the states are allowed by the permanent electric dipole moment of \ce{ND2} ($\mu_b = 1.82(5)$\,D \citealt{brown1979determination}).
		At low temperature, given the energy difference of 15.5\,K between the \emph{ortho} and \emph{para} states, \emph{o}-\ce{ND2} and \emph{p}-\ce{ND2} should be considered as different species.
		
		The network of fine and hyperfine interactions in \ce{ND2} is analogous to that described earlier for \ce{NHD}, while the coupling scheme of angular momentum changes as follows:
		\begin{subequations} \label{eq:coup2}
			\begin{eqnarray} \label{eq:subcoup2}
				&\mathbf{F}_1  =& \mathbf{J}   + \mathbf{I}_\mrm{N}       \,, \\
				&\mathbf{F}    =& \mathbf{F}_1 + \mathbf{I}_\mrm{D,\,tot} \,.
			\end{eqnarray}
		\end{subequations}
		A detailed account of the spectroscopic features of \ce{ND2} can be found in \citet{melosso2017terahertz}.
		
		\section{Partition function values}\label{app:partit}
		
		The spectroscopic parameters from \citet{bizzocchi2020nhd} and \citet{melosso2017terahertz} have been input in the \texttt{SPCAT} subroutine \citep{pickett1991} to evaluate the rotational partition function of \ce{NHD} and \ce{ND2}, respectively. All the energy levels up to $J=20$ and $K_a=15$ have been considered.
		The temperature-dependence of $Q_\mrm{rot}$ has been computed analytically for both species at temperatures between 2.725 and 300\,K.	These values are given in Table~\ref{tab:qrot}.
		
		\begin{table}[h!]
			\centering
			\caption{Rotational partition function values of \ce{NHD} and \ce{ND2} computed at different temperatures.}
			\label{tab:qrot}
			\begin{tabular}{c cc}
				\hline\hline \\[-1ex]
				Temperature (K) & \ce{NHD} & \ce{ND2} \\[0.5ex]
				\hline \\[-1.5ex]
				300.000 &  6420.8713  &  7580.4217  \\[0.5ex]
				225.000 &  4176.0818  &  4927.1928  \\[0.5ex]
				150.000 &  2283.1782  &  2689.7500  \\[0.5ex]
				75.000 &   820.8872  &   962.6221  \\[0.5ex]
				37.500 &   301.1282  &   349.7845  \\[0.5ex]
				18.750 &   115.5024  &   130.1662  \\[0.5ex]
				9.375 &    52.9533  &    48.8603  \\[0.5ex]
				5.000 &    38.1321  &    23.5653  \\[0.5ex]
				2.725 &    36.0453  &    18.3627  \\[0.5ex]
				\hline\hline
			\end{tabular}
		\end{table}

		\section{List of the used HFS components}\label{app:list}
		
		Tables~\ref{tab:list1}--\ref{tab:list4} list the rest frequencies of the observed hyperfine components of \ce{NHD} and \ce{ND2}.
		The LTE intensities (computed at 300\,K and labeled LGINT in the tables) and the HFS quantum numbers are also given.
		Spectral predictions are based on the spectroscopic studies reported in \citet{melosso2017terahertz} and in \citet{bizzocchi2020nhd}.
		
		\begin{table}[htb!]
			\centering
			\footnotesize
			\caption{List of HFS components used to reproduce the $J = \tfrac{3}{2} \leftarrow \tfrac{1}{2}\,,\,N_{Ka,\,Kc} = 1_{1,\,1} \leftarrow 0_{\,0,\,0}$ transition of \ce{NHD}.}
			\label{tab:list1}
			\begin{tabular}{cc ccc ccc}
				\hline\hline \\[-1ex]
				Frequency & LGINT & \multicolumn{3}{c}{Up. state} & \multicolumn{3}{c}{Lo. state} \\[0.5ex]
				(MHz) & (nm$^2$ MHz) & $F_1^\prime$ & $F_2^\prime$ & $F^\prime$ & $F_1$ & $F_2$ & $F$  \\[0.5ex]
				\hline \\[-1.5ex]
				770684.033 &  -3.5757 & 0.5 & 0 & 1  &  1.5 & 1 & 1  \\[0.5ex]
				770697.605 &  -3.4229 & 1.5 & 1 & 2  &  1.5 & 1 & 1  \\[0.5ex]
				770701.070 &  -3.0725 & 1.5 & 1 & 2  &  1.5 & 1 & 2  \\[0.5ex]
				770703.764 &  -3.5532 & 1.5 & 1 & 0  &  1.5 & 1 & 1  \\[0.5ex]
				770706.007 &  -3.2813 & 1.5 & 1 & 1  &  1.5 & 1 & 2  \\[0.5ex]
				770713.045 &  -3.3739 & 1.5 & 2 & 1  &  1.5 & 2 & 1  \\[0.5ex]
				770714.635 &  -3.2184 & 1.5 & 2 & 2  &  1.5 & 2 & 2  \\[0.5ex]
				770717.184 &  -3.0809 & 1.5 & 2 & 3  &  1.5 & 2 & 3  \\[0.5ex]
				770729.939 &  -3.1592 & 0.5 & 1 & 2  &  0.5 & 1 & 2  \\[0.5ex]
				770731.407 &  -2.6209 & 2.5 & 2 & 3  &  1.5 & 1 & 2  \\[0.5ex]
				770732.303 &  -3.3683 & 0.5 & 1 & 1  &  0.5 & 1 & 1  \\[0.5ex]
				770733.192 &  -3.5106 & 0.5 & 1 & 1  &  0.5 & 1 & 0  \\[0.5ex]
				770733.891 &  -2.9068 & 2.5 & 2 & 2  &  1.5 & 1 & 1  \\[0.5ex]
				770735.685 &  -3.4570 & 0.5 & 1 & 0  &  0.5 & 1 & 1  \\[0.5ex]
				770735.961 &  -3.2356 & 2.5 & 2 & 1  &  1.5 & 1 & 0  \\[0.5ex]
				770737.355 &  -3.1949 & 2.5 & 2 & 2  &  1.5 & 1 & 2  \\[0.5ex]
				770737.593 &  -3.2680 & 2.5 & 2 & 1  &  1.5 & 1 & 1  \\[0.5ex]
				770739.631 &  -2.8049 & 0.5 & 1 & 2  &  0.5 & 0 & 1  \\[0.5ex]
				770743.306 &  -2.7096 & 2.5 & 3 & 2  &  1.5 & 2 & 1  \\[0.5ex]
				770744.024 &  -2.5392 & 2.5 & 3 & 3  &  1.5 & 2 & 2  \\[0.5ex]
				770745.755 &  -3.3952 & 0.5 & 1 & 1  &  0.5 & 0 & 1  \\[0.5ex]
				770746.127 &  -2.3691 & 2.5 & 3 & 4  &  1.5 & 2 & 3  \\[0.5ex]
				770746.604 &  -2.6141 & 1.5 & 2 & 3  &  0.5 & 1 & 2  \\[0.5ex]
				770748.263 &  -2.9477 & 1.5 & 2 & 2  &  0.5 & 1 & 1  \\[0.5ex]
				770748.379 &  -3.4022 & 2.5 & 3 & 2  &  1.5 & 2 & 2  \\[0.5ex]
				770751.746 &  -3.4616 & 1.5 & 2 & 1  &  0.5 & 1 & 1  \\[0.5ex]
				770751.990 &  -3.3905 & 2.5 & 3 & 3  &  1.5 & 2 & 3  \\[0.5ex]
				770752.022 &  -3.4349 & 1.5 & 2 & 2  &  0.5 & 1 & 2  \\[0.5ex]
				770752.634 &  -3.3344 & 1.5 & 2 & 1  &  0.5 & 1 & 0  \\[0.5ex]
				770762.836 &  -3.6026 & 1.5 & 1 & 2  &  1.5 & 2 & 3  \\[0.5ex]
				770778.684 &  -3.4439 & 0.5 & 0 & 1  &  0.5 & 1 & 2  \\[0.5ex]
				770792.256 &  -3.4614 & 1.5 & 1 & 2  &  0.5 & 1 & 2  \\[0.5ex]
				770793.173 &  -3.2079 & 2.5 & 2 & 3  &  1.5 & 2 & 3  \\[0.5ex]
				770801.949 &  -3.5909 & 1.5 & 1 & 2  &  0.5 & 0 & 1  \\[0.5ex]
				770806.886 &  -3.6006 & 1.5 & 1 & 1  &  0.5 & 0 & 1  \\[0.5ex]
				\hline\hline
			\end{tabular}
		\end{table}
		
		\begin{table}[htb!]
			\centering
			\footnotesize
			\caption{List of HFS components used to reproduce the $J = \tfrac{1}{2} \leftarrow \tfrac{1}{2}\,,\,N_{Ka,\,Kc} = 1_{1,\,1} \leftarrow 0_{\,0,\,0}$ transition of \ce{NHD}.}
			\label{tab:list2}
			\begin{tabular}{cc ccc ccc}
				\hline\hline \\[-1ex]
				Frequency & LGINT & \multicolumn{3}{c}{Up. state} & \multicolumn{3}{c}{Lo. state} \\[0.5ex]
				(MHz) & (nm$^2$ MHz) & $F_1^\prime$ & $F_2^\prime$ & $F^\prime$ & $F_1$ & $F_2$ & $F$  \\[0.5ex]
				\hline \\[-1.5ex]
				775951.7799 & -3.3743 & 1.5 & 2 & 2 & 1.5 & 1 & 1 \\[0.5ex]
				775954.2373 & -3.0381 & 1.5 & 2 & 3 & 1.5 & 1 & 2 \\[0.5ex]
				775983.7269 & -3.2975 & 1.5 & 1 & 2 & 1.5 & 2 & 3 \\[0.5ex]
				776004.0628 & -3.1888 & 0.5 & 1 & 2 & 1.5 & 1 & 2 \\[0.5ex]
				776004.6459 & -3.3020 & 1.5 & 2 & 1 & 1.5 & 2 & 1 \\[0.5ex]
				776009.0444 & -3.1157 & 1.5 & 2 & 2 & 1.5 & 2 & 2 \\[0.5ex]
				776009.3879 & -3.3118 & 1.5 & 1 & 2 & 0.5 & 1 & 1 \\[0.5ex]
				776009.5205 & -3.5730 & 0.5 & 0 & 1 & 1.5 & 1 & 1 \\[0.5ex]
				776012.9847 & -3.3589 & 0.5 & 0 & 1 & 1.5 & 1 & 2 \\[0.5ex]
				776013.1471 & -3.2111 & 1.5 & 1 & 2 & 0.5 & 1 & 2 \\[0.5ex]
				776013.9374 & -3.5115 & 1.5 & 1 & 1 & 0.5 & 1 & 2 \\[0.5ex]
				776016.0039 & -2.8268 & 1.5 & 2 & 3 & 1.5 & 2 & 3 \\[0.5ex]
				776022.8395 & -3.5879 & 1.5 & 1 & 2 & 0.5 & 0 & 1 \\[0.5ex]
				776023.6298 & -3.3538 & 1.5 & 1 & 1 & 0.5 & 0 & 1 \\[0.5ex]
				776042.6716 & -3.3983 & 1.5 & 2 & 2 & 0.5 & 1 & 1 \\[0.5ex]
				776044.2348 & -3.6053 & 1.5 & 2 & 1 & 0.5 & 1 & 0 \\[0.5ex]
				776045.4241 & -3.1291 & 1.5 & 2 & 3 & 0.5 & 1 & 2 \\[0.5ex]
				776046.4309 & -3.5307 & 1.5 & 2 & 2 & 0.5 & 1 & 2 \\[0.5ex]
				776052.6624 & -3.5309 & 0.5 & 1 & 0 & 1.5 & 2 & 1 \\[0.5ex]
				776057.7495 & -3.1714 & 0.5 & 1 & 1 & 1.5 & 2 & 2 \\[0.5ex]
				776065.8293 & -2.9268 & 0.5 & 1 & 2 & 1.5 & 2 & 3 \\[0.5ex]
				776104.1715 & -3.4331 & 0.5 & 0 & 1 & 0.5 & 1 & 2 \\[0.5ex]
				\hline\hline
			\end{tabular}
		\end{table}
		
		\begin{table}[htb!]
			\centering
			\footnotesize
			\caption{List of HFS components used to reproduce the $J = \tfrac{3}{2} \leftarrow \tfrac{1}{2}\,,\,N_{Ka,\,Kc} = 1_{1,\,1} \leftarrow 0_{\,0,\,0}$ transition of \ce{ND2}.}
			\label{tab:list5}
			\begin{tabular}{cc ccc ccc}
				\hline\hline \\[-1ex]
				Frequency & LGINT & \multicolumn{3}{c}{Up. state} & \multicolumn{3}{c}{Lo. state} \\[0.5ex]
				(MHz) & (nm$^2$ MHz) & $F_1^\prime$ & $I_\mrm{D,\,tot}^\prime$ & $F^\prime$ & $F_1$ & $I_\mrm{D,\,tot}$ & $F$  \\[0.5ex]
				\hline \\[-1.5ex]
				527133.0435 & -4.2749 & 0.5 & 1 & 0.5 & 1.5 & 1 & 0.5 \\[0.5ex]
				527137.3007 & -4.2731 & 0.5 & 1 & 0.5 & 1.5 & 1 & 1.5 \\[0.5ex]
				527138.6811 & -4.2681 & 0.5 & 1 & 1.5 & 1.5 & 1 & 2.5 \\[0.5ex]
				527145.1770 & -4.1674 & 1.5 & 1 & 2.5 & 1.5 & 1 & 1.5 \\[0.5ex]
				527147.6220 & -3.9877 & 1.5 & 1 & 1.5 & 1.5 & 1 & 0.5 \\[0.5ex]
				527151.0397 & -3.8630 & 1.5 & 1 & 0.5 & 1.5 & 1 & 0.5 \\[0.5ex]
				527151.8792 & -3.5628 & 1.5 & 1 & 1.5 & 1.5 & 1 & 1.5 \\[0.5ex]
				527155.0073 & -3.2897 & 1.5 & 1 & 2.5 & 1.5 & 1 & 2.5 \\[0.5ex]
				527155.2969 & -3.8423 & 1.5 & 1 & 0.5 & 1.5 & 1 & 1.5 \\[0.5ex]
				527161.7095 & -3.8657 & 1.5 & 1 & 1.5 & 1.5 & 1 & 2.5 \\[0.5ex]
				527175.2302 & -3.3744 & 0.5 & 1 & 1.5 & 0.5 & 1 & 1.5 \\[0.5ex]
				527179.4875 & -3.2692 & 0.5 & 1 & 1.5 & 0.5 & 1 & 0.5 \\[0.5ex]
				527182.9457 & -2.9117 & 2.5 & 1 & 2.5 & 1.5 & 1 & 1.5 \\[0.5ex]
				527183.5235 & -3.1343 & 2.5 & 1 & 1.5 & 1.5 & 1 & 0.5 \\[0.5ex]
				527183.6802 & -3.4747 & 0.5 & 1 & 0.5 & 0.5 & 1 & 1.5 \\[0.5ex]
				527185.6745 & -2.6845 & 2.5 & 1 & 3.5 & 1.5 & 1 & 2.5 \\[0.5ex]
				527187.7807 & -3.5532 & 2.5 & 1 & 1.5 & 1.5 & 1 & 1.5 \\[0.5ex]
				527187.9375 & -4.1230 & 0.5 & 1 & 0.5 & 0.5 & 1 & 0.5 \\[0.5ex]
				527191.5565 & -3.0134 & 1.5 & 1 & 2.5 & 0.5 & 1 & 1.5 \\[0.5ex]
				527192.7760 & -3.5051 & 2.5 & 1 & 2.5 & 1.5 & 1 & 2.5 \\[0.5ex]
				527198.2587 & -3.5357 & 1.5 & 1 & 1.5 & 0.5 & 1 & 1.5 \\[0.5ex]
				527201.6764 & -4.3765 & 1.5 & 1 & 0.5 & 0.5 & 1 & 1.5 \\[0.5ex]
				527202.5160 & -3.6385 & 1.5 & 1 & 1.5 & 0.5 & 1 & 0.5 \\[0.5ex]
				527205.9337 & -3.7120 & 1.5 & 1 & 0.5 & 0.5 & 1 & 0.5 \\[0.5ex]
				\hline\hline
			\end{tabular}
		\end{table}
		
		\begin{table}[htb!]
			\centering
			\footnotesize
			\caption{List of HFS components used to reproduce the $J = \tfrac{5}{2} \leftarrow \tfrac{3}{2}\,,\,N_{Ka,\,Kc} = 2_{1,\,2} \leftarrow 1_{\,0,\,1}$ transition of \ce{ND2}.}
			\label{tab:list3}
			\begin{tabular}{cc ccc ccc}
				\hline\hline \\[-1ex]
				Frequency & LGINT & \multicolumn{3}{c}{Up. state} & \multicolumn{3}{c}{Lo. state} \\[0.5ex]
				(MHz) & (nm$^2$ MHz) & $F_1^\prime$ & $I_\mrm{D,\,tot}^\prime$ & $F^\prime$ & $F_1$ & $I_\mrm{D,\,tot}$ & $F$  \\[0.5ex]
				\hline \\[-1.5ex]
				784901.6100 & -3.3624 & 2.5 & 2 & 3.5 & 2.5 & 2 & 3.5 \\[0.5ex]
				784901.8686 & -3.3381 & 2.5 & 0 & 2.5 & 2.5 & 0 & 2.5 \\[0.5ex]
				784903.2122 & -3.2352 & 2.5 & 2 & 4.5 & 2.5 & 2 & 4.5 \\[0.5ex]
				784906.4925 & -3.4213 & 1.5 & 2 & 2.5 & 1.5 & 2 & 2.5 \\[0.5ex]
				784906.9759 & -3.2482 & 1.5 & 2 & 3.5 & 1.5 & 2 & 3.5 \\[0.5ex]
				784907.7212 & -3.3377 & 1.5 & 0 & 1.5 & 1.5 & 0 & 1.5 \\[0.5ex]
				784912.0082 & -3.4046 & 1.5 & 2 & 0.5 & 1.5 & 2 & 1.5 \\[0.5ex]
				784913.9865 & -3.4905 & 1.5 & 2 & 1.5 & 1.5 & 2 & 2.5 \\[0.5ex]
				784920.0864 & -2.8823 & 1.5 & 2 & 2.5 & 0.5 & 2 & 1.5 \\[0.5ex]
				784924.5270 & -2.4925 & 1.5 & 2 & 3.5 & 0.5 & 2 & 2.5 \\[0.5ex]
				784925.9631 & -2.8414 & 1.5 & 0 & 1.5 & 0.5 & 0 & 0.5 \\[0.5ex]
				784927.5805 & -2.9946 & 1.5 & 2 & 1.5 & 0.5 & 2 & 1.5 \\[0.5ex]
				784929.8205 & -2.9028 & 2.5 & 2 & 2.5 & 1.5 & 2 & 1.5 \\[0.5ex]
				784929.8382 & -2.6114 & 2.5 & 2 & 3.5 & 1.5 & 2 & 2.5 \\[0.5ex]
				784931.7629 & -2.3779 & 2.5 & 2 & 4.5 & 1.5 & 2 & 3.5 \\[0.5ex]
				784932.0249 & -3.2870 & 2.5 & 2 & 1.5 & 1.5 & 2 & 0.5 \\[0.5ex]
				784932.2158 & -3.3227 & 1.5 & 2 & 0.5 & 0.5 & 2 & 1.5 \\[0.5ex]
				784932.7588 & -2.6164 & 2.5 & 0 & 2.5 & 1.5 & 0 & 1.5 \\[0.5ex]
				784934.3875 & -3.0821 & 2.5 & 2 & 1.5 & 1.5 & 2 & 1.5 \\[0.5ex]
				784934.6498 & -3.0523 & 1.5 & 2 & 2.5 & 0.5 & 2 & 2.5 \\[0.5ex]
				784934.6762 & -3.2289 & 2.5 & 2 & 0.5 & 1.5 & 2 & 0.5 \\[0.5ex]
				784936.4341 & -3.0168 & 2.5 & 2 & 2.5 & 1.5 & 2 & 2.5 \\[0.5ex]
				784937.6522 & -2.5509 & 3.5 & 2 & 3.5 & 2.5 & 2 & 2.5 \\[0.5ex]
				784937.6570 & -2.7232 & 3.5 & 2 & 2.5 & 2.5 & 2 & 1.5 \\[0.5ex]
				784938.3512 & -2.9047 & 3.5 & 2 & 1.5 & 2.5 & 2 & 0.5 \\[0.5ex]
				784938.5498 & -2.3902 & 3.5 & 2 & 4.5 & 2.5 & 2 & 3.5 \\[0.5ex]
				784940.0965 & -2.4157 & 3.5 & 0 & 3.5 & 2.5 & 0 & 2.5 \\[0.5ex]
				784940.4444 & -3.1259 & 2.5 & 2 & 3.5 & 1.5 & 2 & 3.5 \\[0.5ex]
				784940.4531 & -2.2396 & 3.5 & 2 & 5.5 & 2.5 & 2 & 4.5 \\[0.5ex]
				784941.3117 & -3.2175 & 3.5 & 2 & 1.5 & 2.5 & 2 & 1.5 \\[0.5ex]
				784942.8337 & -3.0461 & 3.5 & 2 & 2.5 & 2.5 & 2 & 2.5 \\[0.5ex]
				784945.3078 & -3.0176 & 3.5 & 2 & 3.5 & 2.5 & 2 & 3.5 \\[0.5ex]
				784948.8334 & -3.1526 & 3.5 & 2 & 4.5 & 2.5 & 2 & 4.5 \\[0.5ex]
				\hline\hline
			\end{tabular}
		\end{table}
		
		\begin{table}[htb!]
			\centering
			\footnotesize
			\caption{List of HFS components used to reproduce the $J = \tfrac{3}{2} \leftarrow \tfrac{1}{2}\,,\,N_{Ka,\,Kc} = 2_{1,\,2} \leftarrow 1_{\,0,\,1}$ transition of \ce{ND2}.}
			\label{tab:list4}
			\begin{tabular}{cc ccc ccc}
				\hline\hline \\[-1ex]
				Frequency & LGINT & \multicolumn{3}{c}{Up. state} & \multicolumn{3}{c}{Lo. state} \\[0.5ex]
				(MHz) & (nm$^2$ MHz) & $F_1^\prime$ & $I_\mrm{D,\,tot}^\prime$ & $F^\prime$ & $F_1$ & $I_\mrm{D,\,tot}$ & $F$  \\[0.5ex]
				\hline \\[-1.5ex]
				786826.0110 & -3.4286 & 1.5 & 2 & 0.5 & 0.5 & 2 & 1.5 \\[0.5ex]
				786828.1584 & -3.2317 & 1.5 & 2 & 1.5 & 0.5 & 2 & 1.5 \\[0.5ex]
				786832.1370 & -3.2842 & 1.5 & 2 & 2.5 & 0.5 & 2 & 1.5 \\[0.5ex]
				786832.2338 & -3.0963 & 2.5 & 2 & 3.5 & 1.5 & 2 & 3.5 \\[0.5ex]
				786834.3398 & -2.9932 & 2.5 & 2 & 2.5 & 1.5 & 2 & 2.5 \\[0.5ex]
				786834.9400 & -3.0538 & 2.5 & 2 & 1.5 & 1.5 & 2 & 1.5 \\[0.5ex]
				786835.0841 & -3.2113 & 2.5 & 2 & 0.5 & 1.5 & 2 & 0.5 \\[0.5ex]
				786836.6556 & -3.1210 & 1.5 & 2 & 2.5 & 0.5 & 2 & 2.5 \\[0.5ex]
				786836.7849 & -3.0547 & 1.5 & 0 & 1.5 & 0.5 & 0 & 0.5 \\[0.5ex]
				786836.9956 & -3.2843 & 2.5 & 2 & 1.5 & 1.5 & 2 & 0.5 \\[0.5ex]
				786836.9995 & -2.6241 & 2.5 & 0 & 2.5 & 1.5 & 0 & 1.5 \\[0.5ex]
				786838.1735 & -2.9018 & 2.5 & 2 & 2.5 & 1.5 & 2 & 1.5 \\[0.5ex]
				786838.3241 & -2.3974 & 2.5 & 2 & 4.5 & 1.5 & 2 & 3.5 \\[0.5ex]
				786838.9615 & -2.6275 & 2.5 & 2 & 3.5 & 1.5 & 2 & 2.5 \\[0.5ex]
				786842.7431 & -2.7050 & 1.5 & 2 & 3.5 & 0.5 & 2 & 2.5 \\[0.5ex]
				786864.1688 & -3.4603 & 0.5 & 2 & 1.5 & 0.5 & 2 & 1.5 \\[0.5ex]
				786868.6874 & -3.0107 & 0.5 & 2 & 1.5 & 0.5 & 2 & 2.5 \\[0.5ex]
				786872.8893 & -3.1436 & 0.5 & 0 & 0.5 & 0.5 & 0 & 0.5 \\[0.5ex]
				786873.4617 & -2.8638 & 0.5 & 2 & 2.5 & 0.5 & 2 & 1.5 \\[0.5ex]
				786877.9803 & -3.0524 & 0.5 & 2 & 2.5 & 0.5 & 2 & 2.5 \\[0.5ex]
				786887.3254 & -3.4508 & 1.5 & 2 & 2.5 & 1.5 & 2 & 3.5 \\[0.5ex]
				786890.0744 & -3.3125 & 1.5 & 2 & 1.5 & 1.5 & 2 & 2.5 \\[0.5ex]
				786891.7607 & -3.4671 & 1.5 & 2 & 0.5 & 1.5 & 2 & 1.5 \\[0.5ex]
				786893.4129 & -2.9898 & 1.5 & 2 & 3.5 & 1.5 & 2 & 3.5 \\[0.5ex]
				786893.6763 & -3.1369 & 1.5 & 0 & 1.5 & 1.5 & 0 & 1.5 \\[0.5ex]
				786894.0531 & -3.4158 & 1.5 & 2 & 2.5 & 1.5 & 2 & 2.5 \\[0.5ex]
				786897.8867 & -3.4000 & 1.5 & 2 & 2.5 & 1.5 & 2 & 1.5 \\[0.5ex]
				\hline\hline
			\end{tabular}
		\end{table}

	\end{appendix}
	
\end{document}